\documentclass[a4paper,twocolumn,aps,prl]{revtex4-2}
\pdfoutput=1

\usepackage{amsmath}
\usepackage{amsfonts}
\usepackage{amssymb}
\usepackage[cal=cm]{mathalfa}
\usepackage{graphicx}
\usepackage{tikz}
\usetikzlibrary{decorations.pathreplacing} 
\usetikzlibrary{decorations.markings} 
\usepackage[list=true]{subcaption}
\usepackage[justification=raggedright]{caption} 
\usepackage{braket}
\usepackage{sidecap}
\usepackage{color}
\usepackage{bbm}
\usepackage{nicefrac}
\usepackage{float}
\usepackage{placeins}
\usepackage[utf8]{inputenc}
\usepackage[normalem]{ulem}
\usepackage{nicefrac}
\newcommand{\stkout}[1]{\ifmmode\text{\sout{\ensuremath{#1}}}\else\sout{#1}\fi}
\usepackage{dsfont}
\usepackage{xcolor, soul} 
\usepackage[normalem]{ulem} 
\usepackage{siunitx}
\usepackage[citebordercolor=blue]{hyperref}
\usepackage{multirow}

\begin{document}

\title{\textcolor{black}{Violation of a Leggett--Garg inequality using ideal negative measurements\\ in neutron interferometry}}

\author{Elisabeth Kreuzgruber$^{1}$}
\email{ekreuzgruber@gmail.com} 
\author{Richard Wagner$^{1}$}
\author{Niels Geerits$^{1}$}
\author{Hartmut Lemmel$^{1,2}$}
\author{\textcolor{black}{Stephan Sponar}$^{1}$}
\email{stephan.sponar@tuwien.ac.at}
\affiliation{%
$^1$Atominstitut, TU Wien, Stadionallee 2, 1020 Vienna, Austria \\
$^2$Institut Laue-Langevin, 38000, Grenoble, France}
\date{\today}

\hyphenpenalty=800\relax
\exhyphenpenalty=800\relax
\sloppy
\setlength{\parindent}{0pt}
\noindent

\begin{abstract}
We report on an experiment that demonstrates the violation of a Leggett--Garg inequality (LGI) with neutrons. LGIs have been proposed in order to assess how far the predictions of quantum mechanics defy `macroscopic realism'.  
With LGIs, correlations of measurements performed on a single system at different times are described.
The measured value of $K =1.120\pm0.007$, obtained in a neutron interferometric experiment, is clearly above the limit $K=1$ predicted by macro-realistic theories.
\end{abstract}

\maketitle

\emph{Introduction.}—The question whether measurable quantities of a quantum object have definite values prior to the actual measurement  is a fundamental issue ever since quantum theory has been introduced more than a century ago. Examples include Bell's inequality \cite{Bell64,Bell66}, which sets bounds on correlations between measurement results of space-like separated components of a composite (entangled) system. A violation of Bell's inequality thus demonstrates that certain predictions of quantum mechanics cannot be reproduced by realistic theories, more precisely, by local hidden variable theories (LHVT). Another prime example is found in the Kochen-Specker theorem \cite{Kochen67}, which stresses the incompatibility of quantum mechanics with a larger class of hidden-variable theories, known as noncontextual hidden-variable theories (NCHVTs). Here it is assumed that the result of a measurement of an observable is predetermined and independent of a suitable (previous or simultaneous) measurement of any other compatible (co-measurable or commuting) observable, i.e., the measurement context. While both, Bell's inequality and tests of the Kochen-Specker theorem, require composite or multiple spatially-separated systems Leggett-Garg inequalities (LGIs) \cite{leggett_quantum_1985} study temporal correlations of a single system, therefore they are often referred to as Bell inequalities \emph{in time}.

Violation of a Bell inequality is a direct witness of entanglement - a very specific feature of quantum mechanics. Contrary, in the case of LGIs the violation occurs due to the \emph{coherent superposition} of system states, which is essentially the most fundamental property of quantum mechanics. In other words LGIs quantify coherence in quantum systems and can consequently be seen as a measure or test of \emph{quantumness}. 

Leggett-Garg inequalities were proposed in 1985 \cite{leggett_quantum_1985} in order to assess whether sets of pairs of sequential measurements on a single quantum system can be consistent with an underlying macro-realistic theory \cite{emary_leggettgarg_2014}. Within the framework of a macro-realistic theory a single macroscopic system fulfills the following two assumptions of macrorealism measured at successive times: (A1) at any given time the system is always in only one of its macroscopically distinguishable states, and (A2) the state of the system can be determined in a non-invasive way, meaning, without disturbing the subsequent dynamics of the system. Quantum mechanics predicts the violation of the inequalities  \textcolor{black}{since it contradicts with both assumptions  (A1) and (A2)}. The (quantum) system under observation has to be measured at different times. Correlations that can be derived from sequences of this measurements 
let us formulate the LGI. The result of these correlation measurements either confirm the absence of a realistic description of the system or the impossibility of measuring the system without disturbing it \cite{emary_leggettgarg_2014}. This will also refuse a well-defined pre-existing value of a measurement.
Recent violations of LGI have been observed in various systems, including photonic qubits \cite{Ruskov06,Jordan06,Dressel11,Goggin11}, nuclear spins in a diamond defect center\cite{Waldherr11}, superconducting qubits in terms of transoms \cite{Palacios10} and flux qubits \cite{Knee16}, nuclear magnetic resonance \cite{Athalye11,Souza11}, and spin-bearing phosphorus impurities in silicon \cite{Knee2012}. Proposed schemes for increasing violations of Leggett-Garg inequalities range from action of an environment on a single qubit in terms of generic quantum channels \cite{Emary13} to open many-body systems in the presence of a nonequilibrium \cite{Arenas19}. In a recent paper \cite{Matsumura22} the authors propose to test a violation of the Leggett-Garg inequality due to the gravitational interaction in a hybrid system consisting of a harmonic oscillator and a spatially localized superposed particle \cite{Bose18}, aiming to probe the \emph{quantumness} of gravity \cite{Bose17,Marletto17}.

The violation of an LGI in an interferometric setup has been proposed in literature theoretically for electrons in \cite{emary_leggett-garg_2012}. \textcolor{black}{The requirement of non-invasive measurements from (A2) is realized in most experiments by utilizing the concept of weak measurements, or by introducing an ancilla system, as implemented in \cite{Knee2012}. Note that even a weak measurement in practice can never be completely non-invasive (due to a non-vanishing measurement strength) and the preparation of the ancilla system will also always be imperfect. However, the experimental procedure from \cite{emary_leggett-garg_2012} realizes \emph{ideal negative measurements} in an interferometer experiment in order to fulfill the requirement of non-invasive measurements from (A2) without the need for an ancilla.}

In this Letter, we present a neutron interferometric experiment, demonstrating a violation of the LGI. In our measurement scheme the single system is represented by the neutron's path in an interferometer. A respective observable is defined and measured non-invasively according to the LGI protocol.

\emph{Leggett--Garg inequality.}—For dichotomous variables $Q_i$, accounting for two \emph{macroscopically distinguishable} states, having outcomes $q_i=\pm1$,  the correlation function for measurements at times $t_i$, $t_j$ is given by 
\begin{equation}\label{eq:Correlator}
C_{ij}=\langle Q_i Q_j\rangle=\sum_{q_i q_j=\pm} q_i q_j P\big(q_i(t_i),q_j(t_j)\big),
\end{equation}
where $P(q_i(t_i),q_j(t_j))$ denotes the \emph{joint} probability of obtaining the  measurement results $q_i$ at time $t_i$ and $q_j$ at time $t_j$. 
Considering Eq.(\ref{eq:Correlator}) for three experimental sets with $i,j\in\{1,2,3\}$ yields the LGI 
\begin{equation}\label{eq:lgi}
K \equiv C_{21} + C_{32}-C_{31},
\end{equation}
where $K$ denotes the \emph{Leggett-Garg correlator}, with limits $-3\leq K \leq 1$. Since the three correlators are derived from probabilities with $|C_{ij}|\leq 1$, the lower limit cannot be violated. However, quantum mechanics allows for a violation of the upper bound. In a two-level system, the maximum obtainable violation is $K=1.5$ \cite{emary_leggettgarg_2014}.

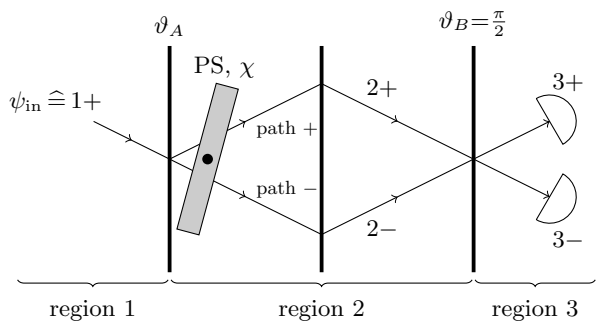
\begin{figure}[h!]
	\centering	
	\begin{tikzpicture}
	\draw[line width=0.5mm] (-2,1.5) -- (-2,-1.5) node[pos=0,above] {$\vartheta_A$};
	\draw[line width=0.5mm] (0,1.5) -- (0,-1.5);
	\draw[line width=0.5mm] (2,1.5) -- (2,-1.5) node[pos=0,above] {$\vartheta_B$=$\frac{\pi}{2}$};;
	\draw[postaction=decorate,decoration={
		markings,
		mark=at position 0.5 with {\arrow{>}}}] (-3,0.5) -- (-2,0) node[left,above,xshift=-1.5cm,yshift=0.5cm] {$\psi_\text{in}\,\widehat{=} \,1+$} ;
	\draw[postaction=decorate,decoration={
		markings,
		mark=at position 0.5 with {\arrow{>}}}] (-2,0) -- (0,-1) node[left,above,xshift=-0.45cm,yshift=0.4cm,scale=0.8] {path $-$};
	\draw[postaction=decorate,decoration={
		markings,
		mark=at position 0.5 with {\arrow{>}}}] (0,-1) -- (2,0)  node[pos=0.4,below,yshift=-0.1cm] {$2-$};
	\draw[postaction=decorate,decoration={
		markings,
		mark=at position 0.5 with {\arrow{>}}}] (-2,0) -- (0,1) node[left,below,xshift=-0.45cm,yshift=-0.4cm,scale=0.8] {path $+$};
	\draw[postaction=decorate,decoration={
		markings,
		mark=at position 0.5 with {\arrow{>}}}] (0,1) -- (2,0)  node[pos=0.4,above,yshift=0.1cm] {$2+$};
	\draw[draw=black,fill=gray!40,rotate around={-15:(-1.5,0.)}]  (-1.65,-1) -- (-1.35,-1) -- (-1.35,1) -- (-1.65,1) -- cycle node[above,xshift=0.85cm,yshift=1.75cm] {PS, $\chi$};
	\fill (-1.5,0.) circle [radius=2pt];
	\draw[postaction=decorate,decoration={
		markings,
		mark=at position 1 with {\arrow{>}}}] (2,0) -- (3,0.5);
	\draw[postaction=decorate,decoration={
		markings,
		mark=at position 1 with {\arrow{>}}}] (2,0) -- (3,-0.5);
	\draw[rotate around={30:(3,0.5)}] (3,0.85) -- (3,0.15) arc[start angle=-90, end angle=90,radius=0.35cm] -- (3,0.85) node[yshift=6,xshift=12] {$3+$};
	\draw[rotate around={-30:(3,-0.5)}] (3,-0.85) -- (3,-0.15) arc[start angle=90, end angle=-90,radius=0.35cm] -- (3,-0.85) node[yshift=-8,xshift=12] {$3-$};
	\draw[decorate,decoration={brace}] (-2.02,-1.6) -- (-4.,-1.6) node[midway,yshift=-0.4cm] {region 1};
	\draw[decorate,decoration={brace}] (1.98,-1.6) -- (-1.98,-1.6) node[midway,yshift=-0.4cm] {region 2};
	\draw[decorate,decoration={brace}] (3.6,-1.6) -- (2.02,-1.6) node[midway,yshift=-0.4cm] {region 3};
	\end{tikzpicture}
	\caption[]{Regions in the Mach-Zehnder interferometer and setup for determination of correlator $C_{31}$.}
	\label{fig:lgi_regions}
\end{figure}

The basic idea behind the experimental procedure as proposed by Emary et al. in \cite{emary_leggett-garg_2012}, is to map the temporal structure (or measurement time $t_i$) of LGI onto real-space coordinates, more precisely onto three distinct regions of the interferometer, indicated by the index $\alpha\in\{1,2,3\}$, cf. Fig. \ref{fig:lgi_regions}. Within each region the two paths of the interferometer constitute a qubit. The measurement of the qubit's state, denoted as $q_i=\pm1$, therefore results in a “which-way” measurement \cite{Englert96} in the particular region of interest. While a click of a detector in e.g. the $+$ arm of region 2 ($q_2=+1$) is a strongly \emph{invasive measurement}, on the other hand the  absence of a detector response implies $q_2=-1$ and does not disturb the system at all. It accounts for the required non-invasive measurement (A2) in terms of an \emph{ideal negative measurement}. 

In our neutron interferometric  realization of \cite{emary_leggett-garg_2012} neutrons enter the IFM via the + port of region 1. Hence, it is not necessary to measure in region 1 and the noninvasive measurability is granted.  The first plate of the IFM consists of a tunable beamsplitter characterized by parameter $\vartheta_A$, which is schematically illustrated in Fig.\,\ref{fig:lgi_regions}. The theoretical maximum of $K=1.5$ is obtained for $\vartheta_A=\vartheta_B=\pi/3$ and phase shift $\chi=0$. However, in our setup with fixed $\vartheta_B=\pi/2$ (usual 50:50 beamsplitter), the maximal possible violation is $K=\sqrt{2}$ (for $\vartheta_A=\pi/4$). 

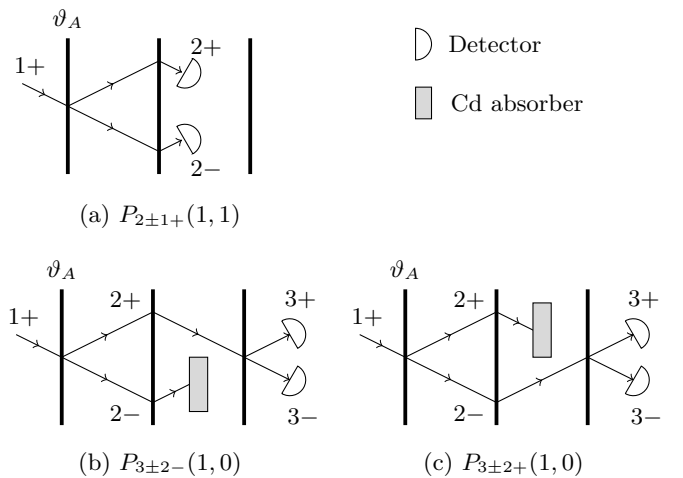
\begin{figure}[!b]
	\vspace*{0.01\textwidth}
	\centering
	\begin{subfigure}[t]{0.49\textwidth}
		\begin{subfigure}[t]{0.48\textwidth}
			\centering
			\begin{tikzpicture}[scale=0.6]
			\draw[line width=0.5mm] (-2,1.5) -- (-2,-1.5) node[pos=0,above] {$\vartheta_A$};
			\draw[line width=0.5mm] (0,1.5) -- (0,-1.5);
			\draw[line width=0.5mm] (2,1.5) -- (2,-1.5) ;
			\draw[postaction=decorate,decoration={
				markings,
				mark=at position 0.5 with {\arrow{>}}}] (-3,0.5) -- (-2,0) node[left,above,xshift=-.5cm,yshift=0.3cm] {$1+$};
			\draw[postaction=decorate,decoration={
				markings,
				mark=at position 0.5 with {\arrow{>}}}] (-2,0) -- (0,-1);
			\draw[postaction=decorate,decoration={
				markings,
				mark=at position 1 with {\arrow{>}}}] (0,-1) -- (0.5,-0.75);
			\draw[postaction=decorate,decoration={
				markings,
				mark=at position 0.5 with {\arrow{>}}}] (-2,0) -- (0,1) ;
			\draw[postaction=decorate,decoration={
				markings,
				mark=at position 1 with {\arrow{>}}}] (0,1) -- (0.5,0.75) ;
			\draw[rotate around={-30:(1,0.5)}] (.5,0.85) -- (.5,0.15) arc[start angle=-90, end angle=90,radius=0.35cm] -- (0.5,0.85) node[yshift=4,xshift=5] {$2+$};
			\draw[rotate around={30:(1,-0.5)}] (.5,-0.85) -- (.5,-0.15) arc[start angle=90, end angle=-90,radius=0.35cm] -- (.5,-0.85) node[yshift=-6,xshift=5] {$2-$};
			\path[rotate around={30:(3,0.5)}] (3,0.85) -- (3,0.15) arc[start angle=-90, end angle=90,radius=0.35cm] -- (3,0.85);
			\end{tikzpicture}
			\caption{$P_{2\pm1+}(1,1)$}
		\end{subfigure}
		\hspace{0.01\textwidth}
		\begin{subfigure}[t]{0.48\textwidth}
			\centering
			\begin{tikzpicture}[scale=0.6]
			\draw (0,1.5) -- (0,0.8) arc[start angle=-90, end angle=90,radius=0.35cm] -- (0,01.5) node[right,xshift=9,yshift=-6] {Detector};
			\draw[draw=black,fill=gray!30] (0,-0.5) rectangle (0.35,0.2) node[right,xshift=4,yshift=-6] {Cd absorber};
			\path (0,-2) rectangle (0.4,-0.6);
			\end{tikzpicture}
		\end{subfigure} 
	\end{subfigure}\\
	\vspace*{0.01\textwidth}
	\begin{subfigure}[t]{0.49\textwidth}
		\begin{subfigure}[t]{0.48\textwidth}
			\centering
			\begin{tikzpicture}[scale=0.6]
			\draw[line width=0.5mm] (-2,1.5) -- (-2,-1.5) node[pos=0,above] {$\vartheta_A$};
			\draw[line width=0.5mm] (0,1.5) -- (0,-1.5);
			\draw[line width=0.5mm] (2,1.5) -- (2,-1.5) ;
			\draw[postaction=decorate,decoration={
				markings,
				mark=at position 0.5 with {\arrow{>}}}] (-3,0.5) -- (-2,0) node[left,above,xshift=-.5cm,yshift=0.3cm] {$1+$};
			\draw[postaction=decorate,decoration={
				markings,
				mark=at position 0.5 with {\arrow{>}}}] (-2,0) -- (0,-1) node[pos=0.7,below,yshift=-0.1cm] {$2-$};
			\draw[postaction=decorate,decoration={
				markings,
				mark=at position 0.5 with {\arrow{>}}}] (0,-1) -- (1,-0.5);
			\draw[postaction=decorate,decoration={
				markings,
				mark=at position 0.5 with {\arrow{>}}}] (-2,0) -- (0,1) node[pos=0.7,above,yshift=0.1cm] {$2+$};
			\draw[postaction=decorate,decoration={
				markings,
				mark=at position 0.5 with {\arrow{>}}}] (0,1) -- (2,0.);
			\draw[postaction=decorate,decoration={
				markings,
				mark=at position 1 with {\arrow{>}}}] (2,0) -- (3,0.5);
			\draw[postaction=decorate,decoration={
				markings,
				mark=at position 1 with {\arrow{>}}}] (2,0) -- (3,-0.5);
			\draw[rotate around={30:(3,0.5)}] (3,0.85) -- (3,0.15) arc[start angle=-90, end angle=90,radius=0.35cm] -- (3,0.85) node[yshift=8,xshift=7] {$3+$};
			\draw[rotate around={-30:(3,-0.5)}] (3,-0.85) -- (3,-0.15) arc[start angle=90, end angle=-90,radius=0.35cm] -- (3,-0.85) node[yshift=-9,xshift=8] {$3-$};
			\draw[draw=black,fill=gray!30]  (1.2,-1.2) -- (0.8,-1.2) -- (0.8,-0) -- (1.2,-0) -- cycle;
			\end{tikzpicture}
			\caption{$P_{3\pm2-}(1,0)$}
		\end{subfigure}
		\hspace{0.01\textwidth}
		\begin{subfigure}[t]{0.48\textwidth}
			\centering
			\begin{tikzpicture}[scale=0.6]
			\draw[line width=0.5mm] (-2,1.5) -- (-2,-1.5) node[pos=0,above] {$\vartheta_A$};
			\draw[line width=0.5mm] (0,1.5) -- (0,-1.5);
			\draw[line width=0.5mm] (2,1.5) -- (2,-1.5) ;
			\draw[postaction=decorate,decoration={
				markings,
				mark=at position 0.5 with {\arrow{>}}}] (-3,0.5) -- (-2,0) node[left,above,xshift=-.5cm,yshift=0.3cm] {$1+$};
			\draw[postaction=decorate,decoration={
				markings,
				mark=at position 0.5 with {\arrow{>}}}] (-2,0) -- (0,-1) node[pos=0.7,below,yshift=-0.1cm] {$2-$};
			\draw[postaction=decorate,decoration={
				markings,
				mark=at position 0.5 with {\arrow{>}}}] (0,-1) -- (2,0);
			\draw[postaction=decorate,decoration={
				markings,
				mark=at position 0.5 with {\arrow{>}}}] (-2,0) -- (0,1) node[pos=0.7,above,yshift=0.1cm] {$2+$};
			\draw[postaction=decorate,decoration={
				markings,
				mark=at position 0.5 with {\arrow{>}}}] (0,1) -- (1,0.5);
			\draw[postaction=decorate,decoration={
				markings,
				mark=at position 1 with {\arrow{>}}}] (2,0) -- (3,0.5);
			\draw[postaction=decorate,decoration={
				markings,
				mark=at position 1 with {\arrow{>}}}] (2,0) -- (3,-0.5);
			\draw[rotate around={30:(3,0.5)}] (3,0.85) -- (3,0.15) arc[start angle=-90, end angle=90,radius=0.35cm] -- (3,0.85) node[yshift=8,xshift=7] {$3+$};
			\draw[rotate around={-30:(3,-0.5)}] (3,-0.85) -- (3,-0.15) arc[start angle=90, end angle=-90,radius=0.35cm] -- (3,-0.85) node[yshift=-9,xshift=8] {$3-$};
			\draw[draw=black,fill=gray!30]  (1.2,1.2) -- (0.8,1.2) -- (0.8,-0) -- (1.2,-0) -- cycle;
			\end{tikzpicture}
			\caption{$P_{3\pm2+}(1,0)$}
		\end{subfigure} 
	\end{subfigure}
	\caption{Setups to determine probabilities $P_{2\pm1+}$ for correlators $C_{21}$ in (a) and $P_{3\pm2\pm}$ for $C_{32}$ in (b),(c).}
	\label{fig:c32}
\end{figure}
\begin{figure}[h!]
\includegraphics[width=0.45\textwidth]{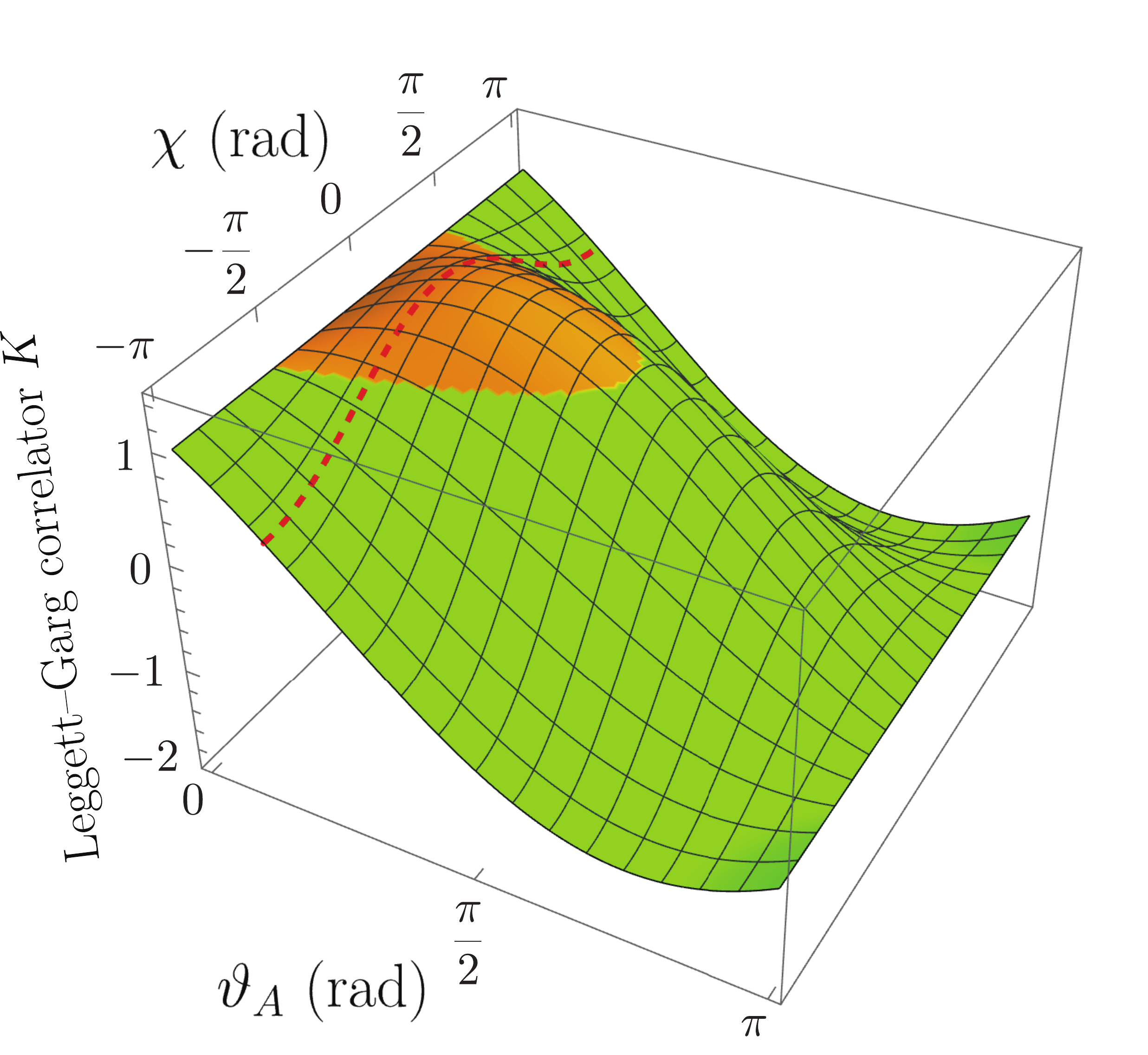}
	\caption{Regions in the parameter space where the LGI can be violated, with fixed $\vartheta_B=\pi/2$. The dashed red line indicates our experimental parameter settings.}	\label{fig:lgi_parspace}
\end{figure}

We define $P_{\alpha\pm,\beta\pm}(n_\alpha,n_\beta)$ as the joint probability that two detectors placed at position $\alpha\pm$ and $\beta\pm$ respectively detect ($n=1$) or don't detect a neutron ($n=0$), where $\alpha$ and $\beta$ specify the region and $\pm$ the path. Then the correlator, as defined in Eq.(\ref{eq:Correlator}), between regions $\alpha$ and $\beta$ is given by
\begin{equation}\label{eq:cab}
 C_{\alpha\beta}=\sum_{q_\alpha,q_\beta=\pm}q_\alpha q_\beta P_{\alpha q_\alpha,\beta q_\beta}(1,1).
\end{equation}
Hence the correlation function for regions 1 and 3, denoted as  $C_{31}$, can simply be expressed as 
$C_{31}=P_{3+,1+}(1,1)-P_{,3-,1+}(1,1)$,
since the neutrons always enter from 1+.
Therefore, the correlation function $C_{31}$ can also be expressed in terms of mariginal probabilities as $C_{31}=P_{3+}(1)-P_{3-}(1)$.
Although not particularly necessary here, 
it is instructive to express $C_{31}$ in terms of \emph{ideal negative} measurements as
\begin{equation}
\begin{split}
C_{31}=&\sum_{q_1,q_3=\pm}q_1 q_3  P_{3 q_\alpha}(1)\big(1-P_{1q_\beta}(0)\big)\\&=-\sum_{q_1,q_3=\pm}q_1 q_3 P_{1q_2,3q_3}(1,0),
\end{split}
\end{equation}
since $P_{1q_1}(0)=1-P_{1q_1}(1)$. A similar expression gives the correlator $C_{21}=P_{1+,2+}(1)-P_{1+,2-}(1)$ which is measured with detectors directly placed in region 2, shown in Fig.~\ref{fig:c32} (a).

For $C_{32}$ all four terms of the sum from Eq.(\ref{eq:cab}) contribute, taking both paths of section 2 into account.
\begin{equation}
 C_{32}=\sum_{q_2,q_3=\pm}q_2 q_3 P_{3q_3,2q_2}(1,1)\\
\end{equation}
Using again $P_{2q_2}(0)=1-P_{2q_2}(1)$ we write the sum as
\begin{equation}\label{eq:c32}
 C_{32}=-\sum_{q_2,q_3=\pm}q_2 q_3 P_{3q_3,2q_2}(1,0)
\end{equation}
in order to account for the non-invasive or ideal negative measurement in section 2. The two pobabilities $P_{3\pm,2-}(1,0)$ are determined by counting the neutrons in path $3+$ and $3-$ respectively under the condition that they have not been counted in pah $2-$. The latter is ensured by placing a beam blocker in path $2-$, cf. Fig.~\ref{fig:c32}(b). The other two pobabilities are measured similarly as shown in Fig.~\ref{fig:c32}(c).

The correlators according to \cite{emary_leggett-garg_2012} for the regions in our setup are calculated as follows
\begin{equation}
\begin{split}
C_{21}=&\cos \vartheta_A\\
C_{32}=& \cos \vartheta_B\\
C_{31}=&\cos \vartheta_A \cos \vartheta_B - \cos \chi \sin \vartheta_A \sin \vartheta_B\\
K=&\cos \vartheta_A+\cos \vartheta_B-\cos \vartheta_A \cos \vartheta_B \\
&+ \cos \chi \sin \vartheta_A \sin \vartheta_B,
\end{split}
\end{equation}
which in our setup, with fixed $\sin\vartheta_B=\frac{\pi}{2}$, $K$ becomes
\begin{equation}
	K=\cos \vartheta_A + \cos \chi \sin \vartheta_A.
\end{equation}

Figure \ref{fig:lgi_parspace} shows the regions in the parameter space ($\vartheta_A$,$\chi$) of our experimental LGI test (with fixed value $\vartheta_B=\pi/2$), where it is in theory possible to violate the LGI with a value $K=\sqrt 2$. $\vartheta_A$ represents the mixing angle of the first interferometer plate, and $\chi$ the phase shifter angle. The resulting $K$ values are shown in green for areas where no violation is possible, and in orange for a possible violation of the LGI. The dashed red line indicates our measurement result in an ideal interferometer.

\emph{Neutron interferometer setup.}—Neutron interferometry \cite{RauchBook, klepp2014fundamental} provides a powerful tool for investigation of fundamental quantum mechanical phenomena. Entanglement between different degrees of freedom (DOF), e.g., the neutron’s spin, path, and energy DOF has been confirmed, and the contextual nature of quantum mechanics has been demonstrated successfully \cite{Sponar21}. In more recent experiments the concept of weak measurements and weak values has been utilized for direct state reconstruction \cite{Denkmayr17}, demonstration of the canonical commutator relation \cite{Wagner21} and studies of which way information \cite{Geppert18,Lemmel2022}.  

\begin{figure}[!b]
	\includegraphics[width=0.49\textwidth]{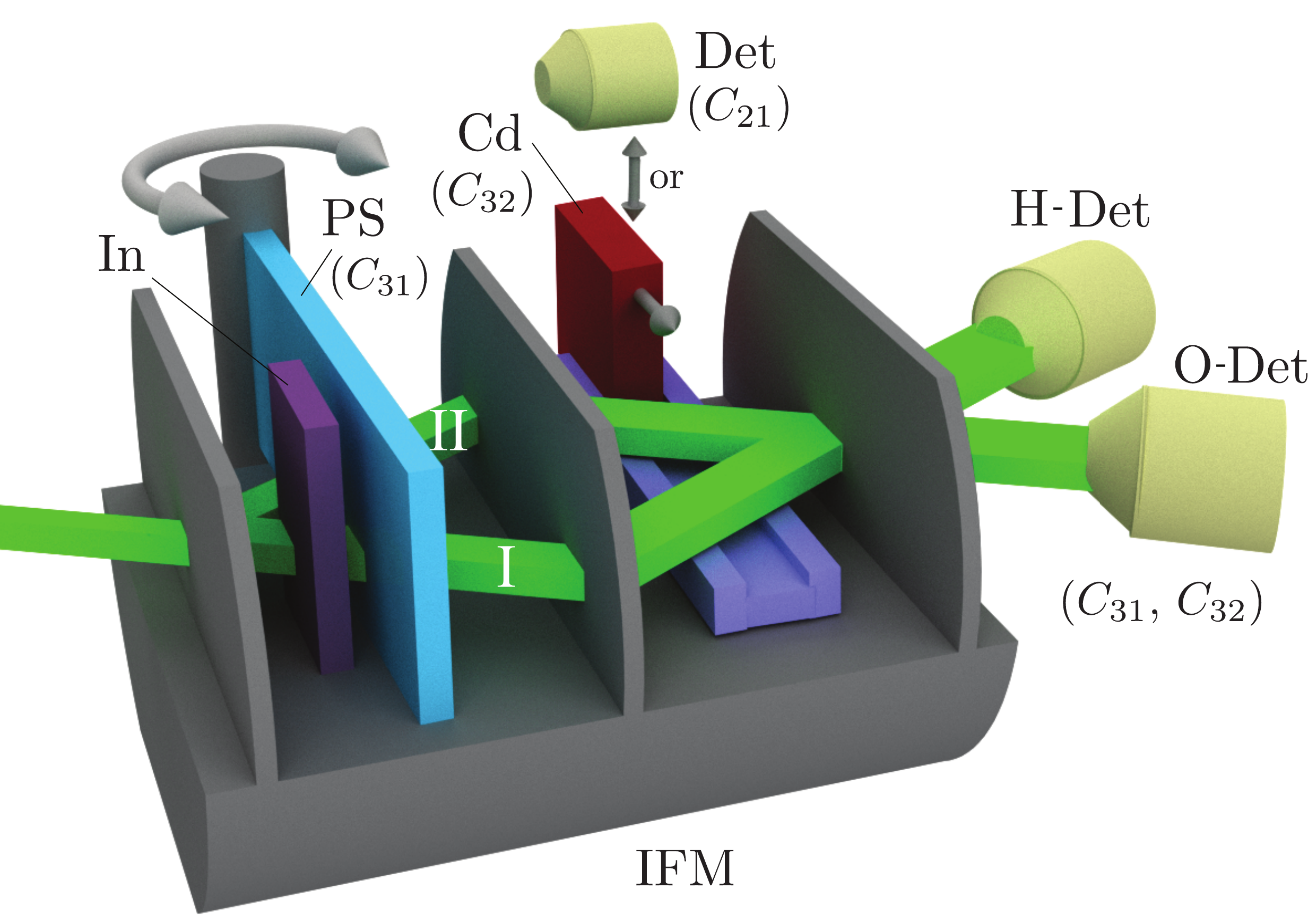}	
 \caption{Unpolarized monochromatic neutrons enter the interferometer and are split into paths I and II. Green indicates the neutron beam, blue the phase shifter and purple the Indium absorber. Detectors O (front) and H (back) as well as the (re)movable detector for $C_{21}$ measurement and Cadmium blocker (red) are shown.}
	\label{fig:meas_scheme}
\end{figure}

The experiment was carried out at the neutron interferometer instrument S18 at the high-flux reactor of the Institute Laue-Langevin (ILL) in Grenoble, France (the experimental data can be found on the ILL data server under \cite{S18data}. A monochromatic unpolarized neutron beam with mean wavelength $\lambda=1.91$\AA \, ($\delta\lambda/\lambda\sim0.02$) and $3 \times 3 \,\mathrm{mm^{2}}$ beam cross section was used to illuminate the interferometer. In order to observe a violation of an LGI in an interferometric experiment, it is necessary to implement a non-50:50 beam splitter at the first plate of the interferometer. This is achieved by placing a partial absorber  behind the first interferometer plate in one of the neutron paths.
The absorber is an Indium slab, about \SI{3}{\milli\metre} thick, placed in path I, resulting in an intensity ratio between paths I and II of about 10:90. The interferometer itself is a symmetric three-plate silicon perfect crystal (triple Laue type), with a plate thickness of \SI{3}{\milli\metre} and a length of \SI{140}{\milli\metre}. A schematic illustration of the interferometric setup is given in Fig.~\ref{fig:meas_scheme}. To obtain interference fringes, a \SI{5}{\milli\metre} Aluminium phase shifter was used. Additional beam blockers for the detection of single path intensities were made of Cadmium. Both the `O' and `H' detectors outside the interferometer and the additional detector for $C_{21}$ measurements were $^3$He proportional counting tubes.

\begin{figure}[!t]
 \includegraphics[width=0.49\textwidth]{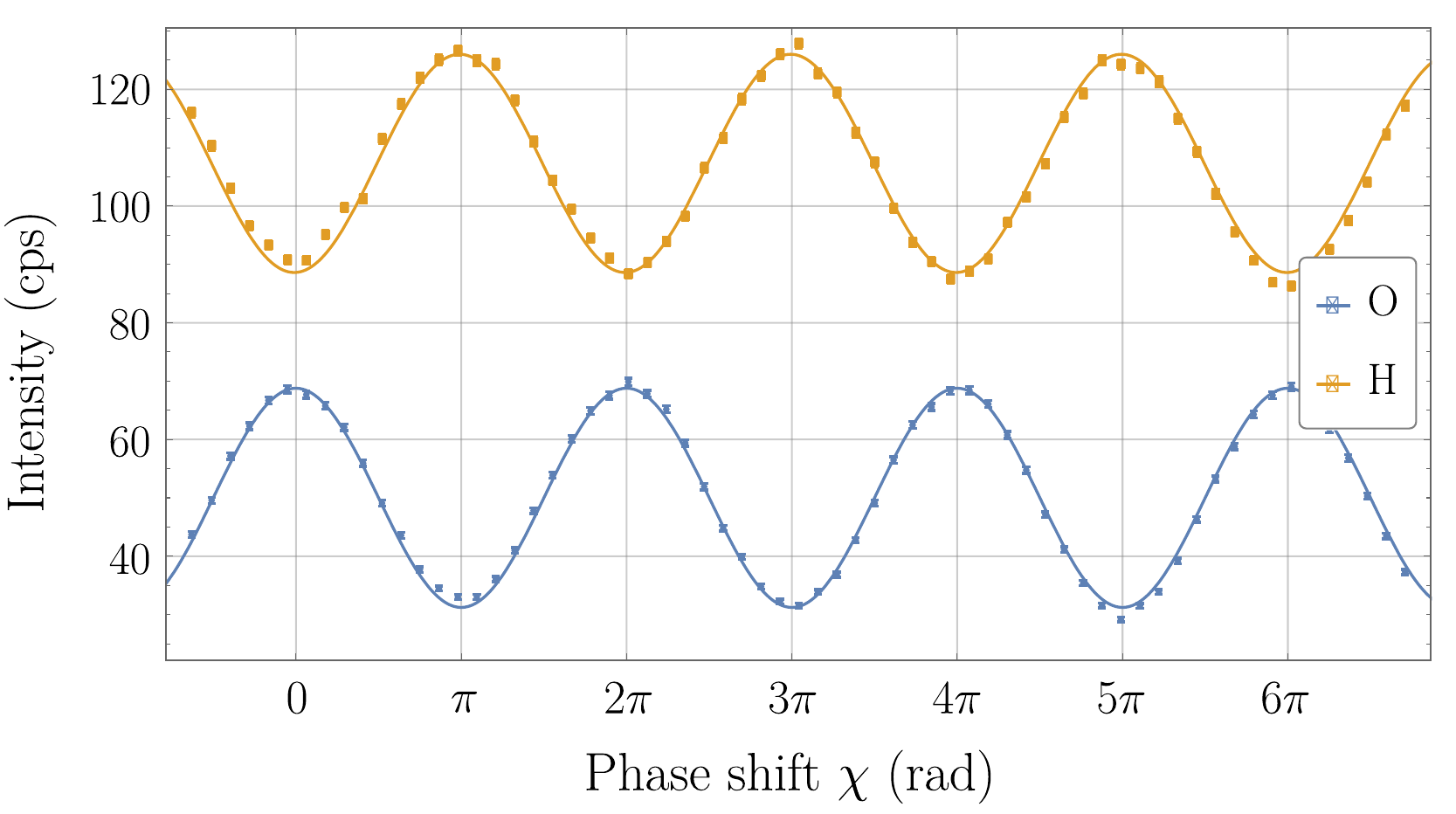}
	\caption{Measurement results for the of $C_{31}$ correlator in terms of interferograms.}
	\label{fig:c31_measurement}
\end{figure}

\begin{figure}[!b]
	\includegraphics[width=0.49\textwidth]{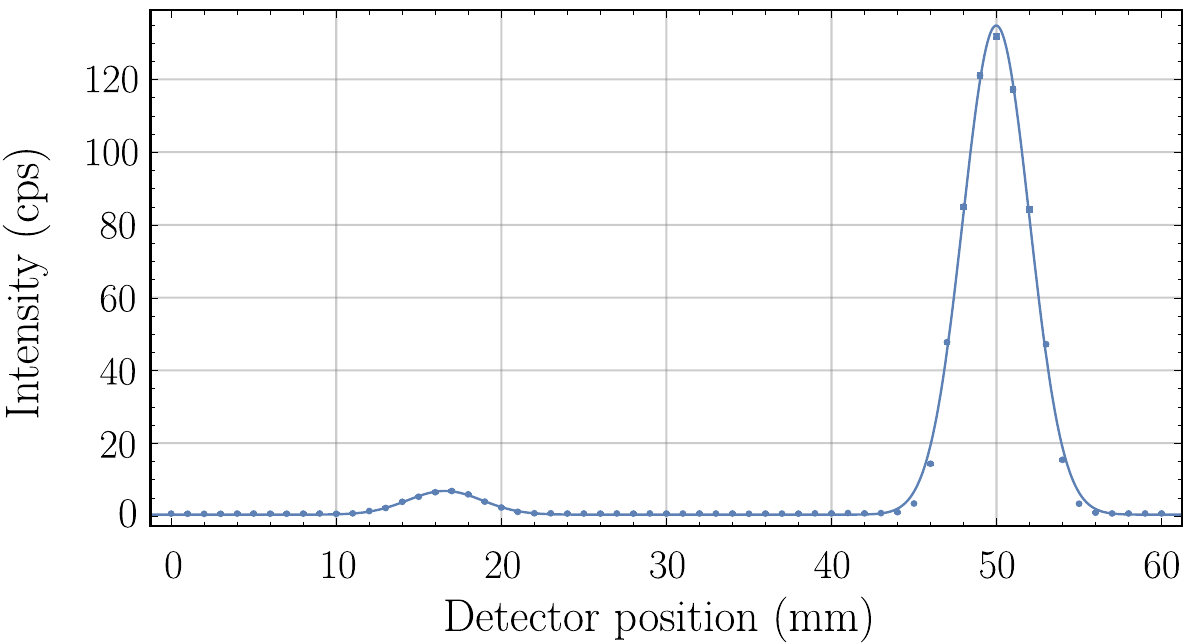}
	\caption{Measurement results for the of $C_{21}$ correlator obtained by transversal scan of movable detector.}
	\label{fig:c21_measurement}
\end{figure}

Determination of correlators $C_{31}$ and $C_{21}$ is straightforward. In both cases it is not necessary to measure non-invasively, since no subsequent measurement on the same state is performed. For $C_{31}$, the measurement is that of a standard interferogram Fig.~\ref{fig:c31_measurement}, \textcolor{black}{with measurement time 180 seconds per phase shifter position}. The correlator $C_{31}$ is calculated via
\begin{equation}
	C_{31}=\frac{N_{3+1+}(\chi)-N_{3-1+}(\chi)}{N_{3+1+}(\chi)+N_{3-1+}(\chi)},
\end{equation}
where $N_{3+1+}(\chi)$ denotes the counts in the H detector and $N_{3-1+}(\chi)$ the counts in the O detector. Due to the cosine behaviour of the recorded interferogram, this correlator is dependent on the position $\chi$ of the phase shifter. For the largest possible violation, the maximum counts in O and minimum in H are used, which corresponds to the position \textcolor{black}{$\chi=2 n \pi$ (where $n\in \mathbb{N}_0$) in Fig.~\ref{fig:c31_measurement}. }

Similarly, the correlator $C_{21}$ is calculated as 
\begin{equation}
	C_{21}=\frac{N_{2+1+}-N_{2-1+}}{N_{2+1+}+N_{2-1+}}
\end{equation}
and is performed as a transversal scan with a pencil-size He-3 detector mounted on a translation stage in region 2 of the interferometer, with measurement time 300 seconds per detector position. Moving first through path I and then through path II, the resulting neutron counts are shown in Fig.~\ref{fig:c21_measurement}, where the separation between both paths is also clearly visible. The $N_{2i1+}$ are the neutron counts in the peak of the respective Gaussian fit to the intensity profiles.

\begin{figure}[!h]
	\includegraphics[width=0.49\textwidth]{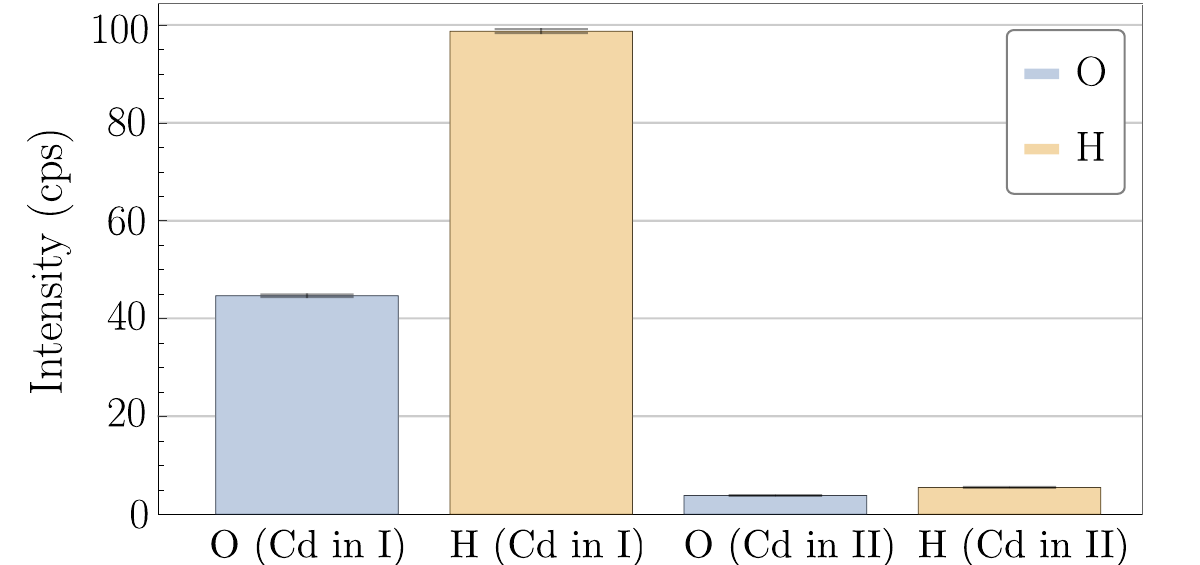}
	\caption{Measurement results for the of $C_{32}$ correlator: neutron counts in detectors O (blue) and H (orange).}
	\label{fig:c32_measurement}
\end{figure}

For correlator $C_{32}$, however, it is crucial to measure non-invasively. This is done by measuring the \textit{absence} of a neutron in a given path due to the Cd blocker, meaning that the neutron has to take the path without the Cd blocker. This is represented by the minus sign in Eq.~\eqref{eq:c32}. Four measurements are performed, with each of the paths blocked in turn and the resulting intensity in detectors O and H recorded for a measurement time of 600 seconds. These results are shown in Fig.~\ref{fig:c32_measurement}. $C_{32}$ becomes
\begin{equation}
	C_{32}=\frac{N_{3+2-}+N_{3-2+}-N_{3+2+}-N_{3-2-}}{N_{3+2-}+N_{3-2+}+N_{3+2+}+N_{3-2-}},
\end{equation}
with $N_{3+2-}$ and $N_{3+2+}$ the neutron counts in the H detector with blocked path II and path I, respectively, and likewise for the O detector in $N_{3-2\pm}$. 
\begin{table}[!t]
\centering
\setlength{\tabcolsep}{5pt}
	\caption{Results of the three correlators $C_{ij}$ and the Leggett--Garg parameter $K$ for violation of the LGI.}
	\sisetup{separate-uncertainty}
	\begin{tabular}{c|c|c|c}
	 $C_{21}$   & $C_{32}$    & $C_{31}$    & $K$     \\  \hline
 $0.903\pm0.002$ & $0.343\pm0.002$ & $0.126\pm0.006$ & $1.120\pm0.007$  \\
	\end{tabular}
	\label{tab:K_results}
\end{table}

\begin{figure}[!h]
	\includegraphics[width=0.49\textwidth]{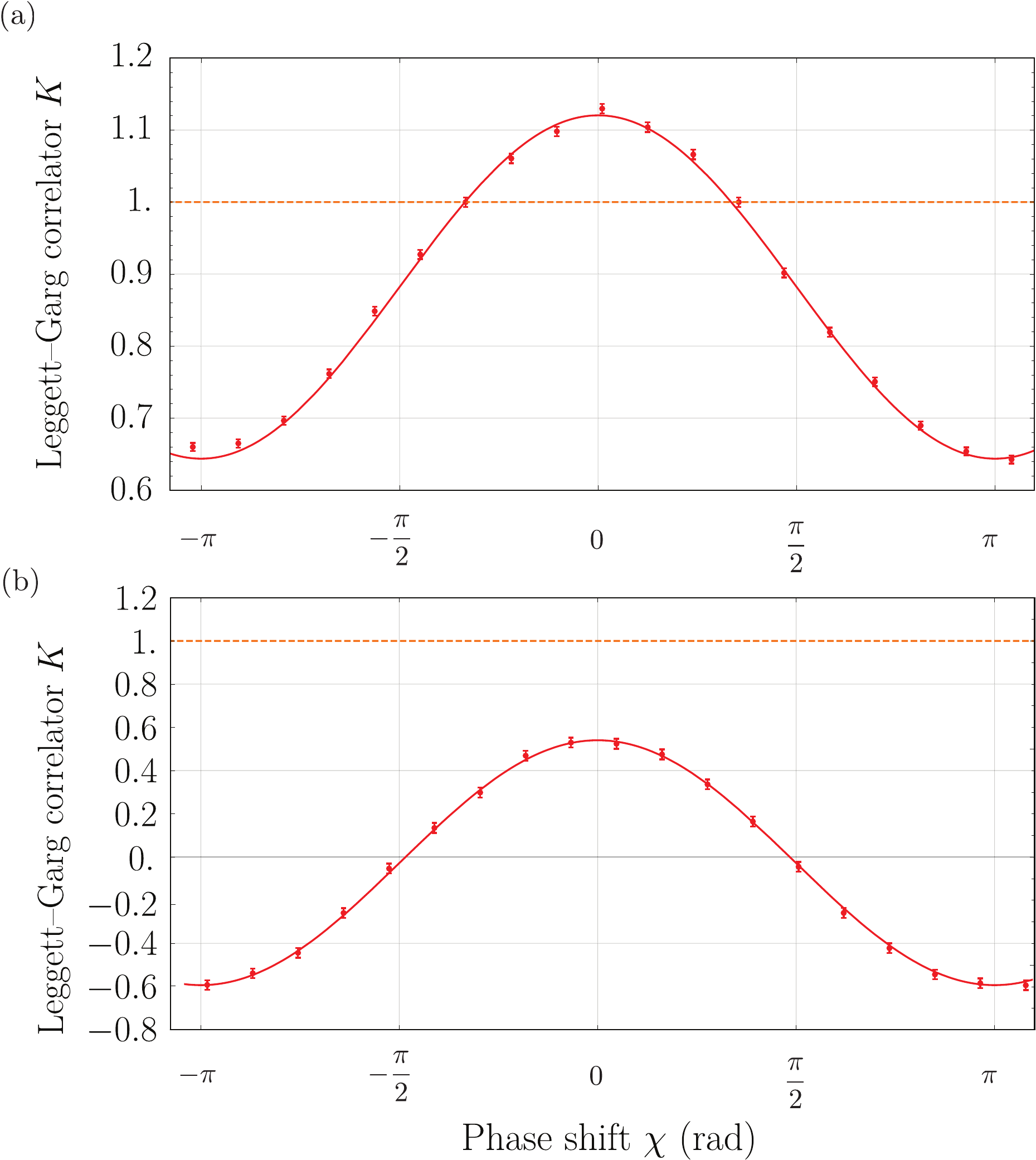}
	\caption{Results of the violation measurements. The dashed line indicates the limit of $K=1$. (a) With Indium absorber a maximal violation is observed at $\chi=0$ and (b) without absorber no violation occurs.}
	\label{fig:k_result}
\end{figure}

\emph{Results.}—In order to demonstrate the experimental violation of the Leggett--Garg inequality, we calculate the correlator $K$, Eq.~\eqref{eq:lgi}. The resulting curve is shown in Fig.~\ref{fig:k_result}, with the maximum at a phase shift of $\chi=0$. With the Indium absorber in path I of the interferometer, a violation of the limit $K=1$ is clearly visible (Fig.~\ref{fig:k_result}(a)). Our results show a significant violation of the LGI by 18 standard deviations $\sigma$ (denoted as $n_\sigma=18$) at the maximum, $K =1.120\pm0.007$. The violation is visible over a wide range of phase shifter values $\chi$. Numeric values of the individual correlators $C_{ij}$ and the final value of $K$ in case of the maximal violation of the LGI are presented in Tab.~\ref{tab:K_results}.
For comparison,  Fig.~\ref{fig:k_result}(b) shows the same measurement procedure  for a symmetric beam splitter ($\vartheta_A=\pi/2$), i.e. without absorber, 
where no violation is possible, resulting in $K=0.540\pm0.023$. 

\emph{Concluding remarks and discussion.}—Our measurement results demonstrate a violation of an LGI by $n_\sigma=18.0$, while the absorberless measurements show no violation. Hence we conclude that neutrons in an interferometer must be understood quantum mechanically. 
An even higher violation can be achieved when the signs in region 3 are switched, and detector O becomes $3+$, detector H $3-$. The correlators $C_{31}$ and $C_{32}$ have to be recalculated accordingly, resulting in $K=1.162\pm0.006$ with $n_\sigma=28$. 
This `additional' violation is due to the  asymmetric nature of the perfect crystal interferometer. Since successive reflections on the crystal lamellas enhance the reflectivity \cite{petrascheck1984} the H detector always receives some phase-independent intensity offset.
The detection loophole is closed due to the high efficiency of our neutron detectors, close to unity. The fair sampling assumption is needed, especially for the correlator $C_{21}$, which is the case for a wide range of experiments of this kind, since simultaneous detection of everything is impossible. 

Finally, we want to emphasize that the interferometric scheme applied in the present work is not limited neutrons, but is in fact completely general and can be used for any quantum particle with nonzero or even zero mass.

\begin{acknowledgements}
	 This work was supported by the Austrian science fund (FWF) Projects No. P 30677 and No. P 34239.
\end{acknowledgements}

\end{document}